# Report on the NSF Workshop on Sustainable Computing for Sustainability (NSF WSCS 2024)

**July 2nd, 2024**

Roch Guérin, Amy McGovern, Klara Nahrstedt
Program co-Chairs

**(Unformatted Final Version)**

# Executive Summary

The workshop brought together approximately 100 in-person participants and a similar, though variable, number of remote participants. Participants came from both industry and academia and, in keeping with the workshop's interdisciplinary focus, spanned a diverse set of disciplines. The workshop's dual theme of "Sustainable Computing" and "Computing for Sustainability" was reflected in a sequence of two similarly structured tracks, which consisted of a panel followed by break-out sessions. They were preceded by a "team science" session intended to expose the participants to the challenges interdisciplinary teams face and offer suggestions to overcome them, with the break-out sessions intended as opportunities to put those into practice.

Panels started with short presentations or position statements by panelists followed by a Q&A session. Panelists and attendees had all been provided ahead of time with a set of questions intended to frame the presentations and discussions. Those questions were distilled from inputs received during an open call that had preceded the workshop. The break-out sessions were structured to be interactive forums to help surface major opportunities and challenges and the resources they would require. Each break-out session had a moderator tasked with guiding the discussions, and "scribes" responsible for documenting those discussions and their findings. Those were presented at the end of the workshop, and are expanded upon in this report. In particular, [Section 5](#) of this report contains the recommendations and findings from the workshop, with a brief synopsis provided next.

**General Recommendations**

***[G1]. Promote open data models and repositories that facilitate data sharing while preserving privacy, and equally important recognize that sustaining the operation of such repositories itself requires dedicated resources.***

This recognizes the importance of data in both making computing more sustainable and in developing solutions to sustainability problems. It also acknowledges the many privacy implications associated with sharing that data, and the need for mechanisms that enable full access without revealing sensitive information. Finally, it highlights the fact that sustaining such repositories over extended periods of time cannot be achieved without access to funding mechanisms to support their operation.

***[G2]. Increase awareness of and accounting for the presence of human and social dimensions in most sustainability problems.***

Human actions have been argued to play a role in many of the factors contributing to sustainability problems, including climate change. However, the focus of this recommendation is in highlighting the fact that human perception and decisions are often instrumental in determining the extent to which solutions that focus on sustainability are eventually adopted. It is, therefore, essential that this be acknowledged and accounted for up-front when developing solutions.

**[G3].** *Sustainability is inherently interdisciplinary and developing effective solutions to sustainability problems requires establishing languages and perspectives that transcend domain boundaries.*

Most sustainability problems involve multiple stakeholders, often with different priorities and a limited understanding of constraints in other domains and the trade-offs they give rise to. Overcoming those challenges calls for "cross-layer" approaches and the creation of programs that are intentional in fostering them.

**Sustainable Computing Recommendations**

*[SC1]. Develop access to funding vehicles that acknowledge the interdisciplinary nature of sustainable computing research and facilitate bringing together the required expertise.*

This is mostly an acknowledgement of the fact that traditional funding programs prioritize a single discipline with the challenges rapidly increasing with the diversity of skills that are required. This is recognized in a number of existing programs but needs to be sustained and possibly expanded to also consider smaller scale efforts that can provide the starting points for larger initiatives.

*[SC2]. Expand collaboration between industry and academia with a focus on technology transfer to increase the odds of impactful outcomes.*

Collaboration between industry and academia is important to grow a strong innovation pipeline in sustainable computing, but ensuring impactful outcomes calls for collaboration models that go beyond traditional research projects. Solutions need to transition from research prototypes to operational systems, and this requires continuity of support over the different phases of the project. The TIP directorate appears to be well-suited to help realize such outcomes.

*[SC3]. Promote sustainable Computing as a Stand-Alone Topic on Both the Educational and Research Fronts.*

Sustainable computing only has a limited presence in today's computer science curriculum, and this needs to change if we are to meet the needs for a workforce with the necessary skills and knowledge. Realizing this goal calls for fostering the creation and sharing of new educational material in that space. Conversely, while there has been work on improving the operational sustainability of datacenters, aspects related to their embodied cost and that of computing in general remain under-studied. Supporting projects that explore this aspect is, therefore, essential to addressing the full sustainability cost of computing.

**Computing for Sustainability Recommendations**

*[C4S1].        Accounting for local impact and conditions in climate models.*

There has been significant progress in global models of the Earth's atmosphere and how and why it is evolving as climate changes. Equally important is the local impact of climate change. There is, therefore, a need for models that can span from the global to the local, and for leveraging our local understanding into actionable outcomes. The latter should be coupled with community-building initiatives that will in turn require dedicated support mechanisms.

### [C4S2].    *Sustainable agriculture as a climate strategy.*

The scale of food production is such that it alone is arguably a potential factor behind climate change. As a result, while investigating the impact of climate on agriculture remains essential, this should be carried out as part of a broader initiative that also explores how food production goals may affect climate change towards developing potential synergies. Realizing such a goal will require encouraging greater communication between the different communities involved.

### [C4S3].    *Harnessing machine learning for more efficient buildings.*

Commercial and residential buildings are major contributors to our energy consumption. Improving their efficiency can, therefore, yield significant sustainability gains. This is, however, challenging because, while data is available, it is often fragmented and proprietary. This, combined with the operational diversity of buildings, often prevents the development of efficient solutions. Learning solutions trained on data produced by individual buildings might offer a promising approach.

# 1.0 Introduction

The workshop on "Sustainable Computing for Sustainability" whose findings are summarized in this document had its origin in a December 2022 report from the CISE Advisory Committee titled "Computing for the Environment" [1]. That report detailed the many opportunities that exist for computing to help tackle the severe environmental challenges we are facing. One of the report's recommendations (Thrust 1) was the pursuit of Multidisciplinary Community Building initiatives, of which this workshop is an example.

Specifically, the CISE AC's report motivated the submission of a collaborative conference proposal[1] titled "Sustainable Computing for Sustainability" proposing one such community building initiative in the form of a workshop to explore the following three premises:

1. That computing can play a major role in tackling modern sustainability challenges, especially those connected to climate change;
2. That computing itself increasingly contributes to sustainability challenges, so that improving its own sustainability is essential;
3. That effective solutions to sustainability challenges will require interdisciplinary collaborations if they are to achieve any form of success.

The proposal was funded (as NSF grants CCF 2334853/4/5), which put in motion the efforts that eventually resulted in this workshop. Following the proposal's title, the goals were to (i) advance the development of research initiatives along the themes of both "sustainable computing" and "computing for sustainability", while also (ii) helping develop and sustain the interdisciplinary teams those initiatives would need. The approach followed to ultimately realize these goals proceeded in three phases:

1. An open solicitation phase, in the form of a "Request for Inputs" (RFI) addressed to both the computing community and communities of computing users, and asking them for inputs to help refine the workshop's themes.
2. A call for contributions targeting the themes identified in the RFI and asking individuals interested in attending the workshop to articulate how they could contribute to advancing those themes.
3. The workshop itself, which was structured as an interactive forum intent on crystallizing major research themes and associated challenges, and the resources needed to tackle them.

The RFI was issued in October 2023 and distributed broadly. It echoed the three premises of the proposal, namely,

1. **Computing for sustainability**: How to harness computing to tackle sustainability problems such as climate change, or redesign the power grid to better handle distributed renewable energy sources?
2. **Sustainable computing**: How do we ensure that computing does not itself contribute to creating sustainability problems and that it accounts for sustainability in its development and operation?
3. **Interdisciplinary teams**. How to best stand-up and sustain the type of interdisciplinary teams required to successfully tackle computing for sustainability and sustainable computing problems?

---

[1] With Roch Guerin, Amy McGovern, and Klara Nahrstedt as co-PIs.

Responders were asked to explicitly identify individual challenges in those areas, articulate their importance and/or uniqueness, identify existing efforts that may be targeting them, and discuss the type of resources required to successfully address them.

The RFI generated over 160 responses. Those responses were reviewed by the three PIs assisted by a Steering Committee and used to refine the *technical themes* the workshop would eventually target. On the technical front, a total of six (6) sub-themes emerged, three (3) in each of the "Sustainable Computing" and "Computing for Sustainability" main themes, as shown in the table below.

| **Sustainable Computing** | | | **Computing for Sustainability** | | |
|---|---|---|---|---|---|
| Integration of computing and smart grids | Modularity & lifecycle of computing | Datacenters, energy & optimization | Computing & climate modeling | Computing for agriculture & food systems | Computing for smart infrastructure & materials |

*Table 1. Technical themes that emerged from the responses to the RFI.*

Responses targeting challenges associated with standing-up and maintaining interdisciplinary teams confirmed that these were multi-faceted and a common obstacle that many researchers faced, and therefore also a topic worth focusing on. Overall, the responses received to the RFI helped shape the next phase and the structure that was chosen for the workshop.

Specifically, the workshop was constructed around three main sessions of equal durations. The first session was devoted to "team science" and organized and led by a team from the [Institute for Research in the Social Sciences (IRISS)](#) at Colorado State University. It was intended to increase participants' awareness of the challenges faced by interdisciplinary teams and how they arise, and help them develop skills and approaches to overcome those challenges (see [Section 2.0](#) for details). Those skills were then expected to be put in practice during the other two sessions on "Sustainable Computing" (see [Section 3.0](#)) and "Computing for Sustainability" (see [Section 4.0](#)), which both consisted of a level-setting plenary in the form of a panel, followed by break-out sessions. The panels were intended to provide initial directions and questions that the break-out sessions would then explore further. A list of "starter" questions were provided (see [Appendix A.1](#)) to help frame the panels and seed the discussions they were expected to trigger. These questions, like the workshop's themes, were extracted from the responses to the RFI and aligned with the themes of [Table 1](#). The break-out sessions that followed the two panels were each mapped to one of the themes of [Table 1](#), with participants assigned ahead of time to different break-out sessions based on a poll of their preferences.

The format of the workshop itself was chosen to be hybrid, *i.e.,* involving both in-person and remote attendees, with the latter participating over zoom. Both in-person and remote participants joined during the plenary components of the workshop, but maintained separate break-out sessions. Each break-out session, whether in-person or remote, had a moderator and several scribes. The moderators were tasked with steering the discussions based on guidelines reviewed during the "[team science](#)" session. The scribes were responsible for taking notes documenting the discussions, and, more importantly, the conclusions that were expected to emerge from those discussions.

The selection of a hybrid format was motivated by the desire to facilitate broad participation, while acknowledging the physical space constraints that limited the number of in-person attendees that could be accommodated.  The selection of in-person attendees was through a "Call for Submissions" that echoed the themes of Table 1 and asked interested individuals to submit a two-page statement of interest.  In their statement of interest, participants were expected to select one of the workshop's themes and offer a perspective on potential contributions to refining the theme and the challenges it might expect to face.  Multiple submissions were allowed for those interested in multiple themes.

The call for submissions generated 150 submissions that were used to select approximately 100 in-person participants.  The selection process took into account expected contributions, as reflected in the submissions, as well as diversity of background, perspective, and expertise towards fostering the community-building goal of the workshop and the interdisciplinary nature of the problems it targeted.  In-person participants also included 11 panelists who were selected from among authors of accepted submissions and by identifying additional individuals whose background and/or experience could contribute to the workshop's goals.  Overall, the registered in-person attendees spanned 26 distinct states and three countries outside the US (Canada, Greece, and the UK), counted representatives from academia, industry and not for profit organizations, and 30% of the attendees identified as women.  In addition, represented academic institutions included a mix of public and private institutions and covered different sizes and locations ranging from large cities to rural locales. In addition to in-person participants, over 150 individuals registered for remote participation.

The rest of this report reviews the discussions that took place during the workshop' sessions and summarizes its main conclusions and findings.  Section 2.0 offers a brief overview of the material presented during the "Team Science" session, and some of the guidelines that were highlighted towards facilitating the success of interdisciplinary teams.  Sections 3.0 and 4.0 capture the discussions and conclusions of the panels and break-out sessions associated with the general themes of "Sustainable Computing" and "Computing for Sustainability."  Section 5.0 summarizes the main findings and recommendations that came out of the workshop, including those pointing to the need for additional resources towards tackling the research challenges it identified.

Structurally, the panels consisted of a sequence of short presentations by each panelist.  These were then followed by a Q&A period to either expand on the themes presented by the panelists or explore new questions of relevance, including those previously disseminated to attendees and panelists (see Appendix A.1).  Break-out sessions varied in their structure and how they organized discussions and summarized their findings. They typically started with an "exploration" period intended to surface questions and challenges related to the session's theme and discussions from the panel.  This was then followed by a "synthesis" period that sought to condense the results of the exploration phase into a small set of major problems and challenges that were viewed as particularly promising or important.  The format of those findings, however, varied across sessions, with some focused on identifying and motivating specific problem areas, while others more explicitly distinguishing between opportunities, challenges, and the skills and tools required to tackle them.Those differences in format are reflected in how the findings of each break-out session are reported, each structured somewhat differently, with Section 5.0 (Summary and Recommendations) offering a unifying synthesis.

The workshop was held on April 16-17 at the NSF facility in Alexandria, VA.  Recordings of the plenary sessions and copies of the slides presented during the panels are available on the [workshop's website](), including a [presentation]() by [Damian Dechev]() from NSF who offered a cross-agency overview of programs of potential relevance to the workshop's themes.

# 2.0 Team Science Session & Recommendations

Producing good science is part expertise and part effective teaming. The [NSF Workshop on Sustainable Computing for Sustainability](#) included a session intended to prepare researchers from various disciplines and organizations for effective teaming using evidence-based strategies. Team science strategies were described and applied at the workshop, and participants were also encouraged to use these teaming practices in their careers as scientists on interdisciplinary projects.

**1. Psychological Safety and Social Sensitivity**

Psychological safety and social sensitivity are crucial elements for effective teaming. They lay the foundation for an environment where team members can take risks, voice their thoughts, and share their opinions without fear of judgment or reprisal. This safety net fosters creativity and open communication, which are key to innovative problem-solving. The workshop underscored the significance of active listening, respect for all contributions, and non-judgmental feedback mechanisms in establishing and maintaining this environment. Social sensitivity refers to a team member's ability to understand and respond to the cues and needs of their colleagues. It requires an awareness of verbal and nonverbal communication and the emotional states of others. Social sensitivity can be enhanced through activities designed to improve emotional intelligence, such as role-playing scenarios or sensitivity training sessions. The workshop used exercises to help participants practice empathy and active listening, critical skills for fostering mutual respect.

**2. Overcoming Disciplinary, Institutional, and Cultural Barriers**

The workshop recognized teams' inherent challenges when members come from diverse academic disciplines or professional backgrounds. Each discipline has its language, methodologies, and research paradigms, which can pose communication barriers. Institutional cultures and norms can vary significantly, influencing work styles, decision-making processes, and expectations. Cultural differences can impact communication styles and conflict-resolution strategies, adding another layer of complexity to the collaboration. The workshop focused on overcoming these barriers by creating a "shared language." This doesn't just refer to terminology but also to shared goals, values, and frameworks that transcend individual disciplinary and cultural backgrounds. Tips provided in the workshop might include regular integration meetings, co-creation sessions for project goals, and joint publications or presentations that help solidify a common understanding and approach.

**3. Setting Expectations**

Team charters are essential documents that outline the team's mission, objectives, roles, and responsibilities. They serve as a reference point that all team members agree upon, helping to prevent misunderstandings. By discussing and agreeing on a team charter, members establish a clear framework for collaboration. This framework can include communication norms, meeting schedules, decision-making protocols, and conflict-resolution strategies, all of which contribute to a harmonious and productive team environment. Recognizing that incentives can vary significantly across different contexts, such as academic versus corporate environments or among people with different disciplinary backgrounds, is crucial. Therefore, it is important to create an environment where motivation and satisfaction within the team are maintained. The workshop discussed aligning team members' personal and professional goals with the project objectives. This might involve adaptable reward systems, recognition of different contributions, and flexible roles catering to individual strengths and career aspirations.

These topics addressed during the NSF Workshop on Sustainable Computing for Sustainability are vital for creating a cohesive and productive team environment, especially in settings that require innovation and cross-disciplinary collaboration. The workshop's focus on these areas sought to share the foundational structures for a comprehensive approach to building effective and resilient teams.

# 3.0 Sustainable Computing

The sustainable computing topic discussion started with a panel that included 6 panelists, [Andrew Chien](#) from the University of Chicago, [Tamar Eilam](#) from IBM Research, [Akshaya Jha](#) from CMU, [Kieran Levin](#) from Framework Inc., [Christine Ortiz](#) from MIT, and [Carole-Jean Wu](#) from Meta. The panelists set the stage for the break-out session in all three areas: (1) Modularity and lifecycle of computing, (2) Data centers, energy and optimization, and (3) Integration of computing with the Grid. Slides from some of the panelists are available on the workshop website ([NSF-WSCS 2024 (edas.info)](#)).

The panelists brought different topics and viewpoints in the areas of modularity and life cycle of computing, data centers and energy, and the integration of computing with the Grid. Specifically, the following themes emerged from the panelists' presentations:

- *Non-modular hardware design* of current computing devices (*e.g.,* laptops), produced via non-reversible assembly processes, causes that they are often unrepairable and storage and RAM are no longer upgradeable. For example, a display lasts up to 10 years where the main board has 2-5 years lifespan. If the hardware would be more modular, the lifespan of a laptop would be much longer, helping greatly with the sustainability of the device.
- *High churn in the supply chain* increases the environmental carbon footprint of devices because many consumer electronics devices are only manufactured for 6 months to 2 years and after that, the supply chain is disassembled, thrown away or recycled. Hence, the repair and replacement of computing components become difficult.
- *Mismatches between software components upgrades and between software and hardware lifecycles* cause that devices are often retired or put offline because software support has ended even though the hardware has significant usable lifespan left. For example, if an operating system cannot not be upgraded due to some OS/hardware dependencies, new security patches cannot be installed and the device will be put offline. New solutions are needed to resolve these mismatches.
- *Existing business models* clash with modularity and prolonging life cycle goals of computing devices since longer lifetime of devices means the decrease of revenues for companies building computing devices (e.g., consumer electronics). Hence, potential incentives and regulations might need to be in place to change the business models.
- *Computer technology materials* such as polymers are developed over decades with specialized properties such as temperature stability, non-flammability and longevity, causing e-waste that does not decompose and hence leads to environmental and societal damage. Hence, the need for new materials for computing technology emerges and one needs to consider bio-inspired materials.
- *AI workloads, exponential growth of data and model sizes and infrastructures* cause *enormous increase of electricity and carbon footprint in data centers.* One has to especially consider the *embodied carbon footprint* which manifests itself through supply chain manufacturing of hardware, racks, and data centers constructions themselves. Furthermore, *operational carbon footprint* must be considered coming from offline AI training, inference phases, and using universal large language models. Hence, we need to understand the interrelationships and tradeoffs of all components in the data center lifecycle. This includes changes in hardware manufacturing, and software designs, developing energy-aware AI models, building high quality models that are not as big, better datasets, reuse of models and different management of SLOs.
- *Power grid will not be decarbonized anytime soon* since the progress of renewable energy sources deployment is slower than anticipated. Furthermore, the curtailment and negative priced (stranded) power are a systemic problem for grid decarbonization. Hence, one should consider computing that could run on excess renewable (stranded) power, one could then achieve zero-carbon cloud with zero marginal carbon power and help the Grid decarbonize.
- *Integration of data centers with power grid* often clashes because these two systems have divergent goals and competing optimization criteria. Hence, a holistic framework must be considered.
- *Placement of power plants close to data centers r*educes the loss of electricity transmission and satisfies the high data centers' electricity demands, but may cause environmental and societal harm in the form of local air pollution, and impact on health and productivity. In contrast, collocation of a wind or solar farm with a data center could help on both fronts, including from the environmental perspective.

The panel concluded with questions from the audience and discussions with the panelists.

This was then followed by parallel break-out sessions along the themes identified from the response to the RFI (See Table 1). Two break-out sessions targeted the themes of "Integration of computing with the (smart) grid" and "Modularity & lifecycle of computing", while the other two both focused on the topic of "Datacenter, energy and optimization" to account for the greater number of attendees who had expressed interest in that topic.

## 3.1 Modularity and Lifecycle of Computing

This break-out session within the sustainable computing track was asked to identify and discuss questions, challenges, approaches and recommendations on how to achieve modularity and longer lifecycle of computing components.

**Questions**: Four important topical questions were discussed.

- What are the *data protocols for sharing supply chain lifecycle* and how do we *extract the environmental impac*t (carbon emission) when manufacturing computing components (embodied carbon) and when running workloads on computing components (operational carbon)? How can we better characterize the cost of production of computing components and properly attribute and perform carbon accounting for designing, reusing, repurposing, and recycling hardware?

- How do we get *retired systems to be reused*? Often, retired systems go to other countries which might not have compute resources. How do we train people in these new regions to utilize these retired systems? Can we consider reusing retired systems as home data centers? Do we need software infrastructure to allow old devices to continue working reliably?
- How do we *fight obsolescence of computing components*? To solve this issue we will need a cultural shift, having major societal commitment to provide, e.g., free repairs for the lifetime of a product. We will need to consider changes in policies and education, identify good second use of systems, and coordinate both software and hardware aspects of planned obsolesense.
- How do we enable *modularity of computing devices*? Modularity is good to increase the lifetime of a device, however there are very little incentives to do so. How do we quantify modularity versus integrated systems tradeoffs in terms of performance, cost, lifetime, environmental cost? What should be the incentives for industry to adopt modular strategies? To solve the modularity issue, one will need to require policy changes, and modularity in hardware and software. For example, many ML models are monolithic per application. Would modularization of ML models bring down the carbon footprint?

**Challenges:** The discussed challenges ranged from technical and standardization challenges to policy and societal challenges. The following technical challenges were identified:

A. Proprietary information makes it extremely difficult to model lifecycle processes;
B. Modularity adds more attack vectors, hence a challenge when reusing electronics is how to protect both sellers and buyers from security exposures and privacy breaches;
C. It is challenging to achieve software and hardware updates with backward compatibility;
D. There is a need for operating system and application modularity designs that would tolerate graceful failures;
E. Hardware challenges include needs for fully open and documented hardware to enable extended lifetime of the hardware, and needs to reevaluate specialized hardware regarding their lifetime and cost.
F. Metrics for sustainable computing and their validation models need to be defined to enable measurements of embodied carbon and operational carbon.
G. Customization is a general trend in computing, but if we want to repurpose computing components, which requires programmability, this is often in opposition to customization. Hence, the challenge is how to repurpose and prolong the lifetime of customized computing components.

Standardization challenges come from situations when new use-cases need to be considered. Standards are great for extensibility, but when new use-cases of computing are needed, one has to often rebuild the infrastructure which means additional cost.

<u>Societal challenges</u> included discussions about

   A. *Governance of recycling,* and overall lifecycle of computing components. Because of governance differences in states, countries, and large corporate industries, it is challenging to apply best practices.
   B. *Customers and obstacles to them making informed decisions* about their computing devices. Because of the overall lack of transparency in the life cycle of their computing devices customers cannot make informed decisions about their computing devices.
   C. Needs for increased awareness and recognition of *sustainable computing research and education in academia.* This increased awareness and recognition includes needs for an increase of sustainability-focused workshops and conferences (and publication venues in general).
   D. *Policy challenges.* We need policy changes to address the current lack of incentives to move sustainable designs forward.

**Approaches, Tools and Recommendations:** The break-out session participants discussed the needs for tools that would incorporate sustainability directly into the design of computing components, tools that would more closely understand the energy consumption at all stages of the manufacturing process and hence provide careful embodied carbon analysis, and tools and data standards that would support the reporting of sustainability metrics in terms of their social cost and related environmental impact.

The following approaches and recommendations were discussed:

- Strong backward compatibility of software and hardware to enable modularity,
- Upgradable motherboards for new generation of devices,
- Thermodynamic considerations from physics to algorithms and co-design across all abstraction layers from physics to the application,
- Leveraging historians and historical case studies to inform future R&D efforts,
- Steering university education towards cross-disciplinary training, conversations and guidelines regarding sustainable computing,
- Productive collaborations of computer and materials scientists towards new materials and fabrication techniques of emerging computing devices,
- Establishing technical and policy standards that would focus on sustainability, and
- Need for stronger academic institution recognition of publishing results in sustainable computing related to the lifecycle of computing.

## 3.2 Data Centers, Energy and Optimization

The break-out session researchers discussed two questions related to data centers, energy and optimization, and several challenges, approaches and recommendations to move forward in this area.

**Questions**:

- The main question discussed was what needs to be done to optimize energy of data centers and what would be the metrics? The answers to this question stressed that we should look beyond minimizing carbon footprint since it does not represent the full sustainability impact. For example, optimization of data centers should take into account the water footprint and the locations (environmental impact) of where the data centers will reside. This means that the current view on data centers regarding carbon emission is oversimplified. One also needs to consider geographical load balancing and balance environmental costs across regions. This may, in particular, lead to co-locating data centers with (renewable) energy sources.
- If one wants to optimize data centers and deliver sustainable data centers, what is the *definition of sustainability*?
    - To derive the definition of sustainability, the break-out session researchers discussed three aspects of computing: *energy usage and management*; *materials usage and recycling*; *water footprint and water recycling*. Interesting data was presented that only 10% of material from data centers is recycled, and 90% of material is going to landfills. Current solutions associate sustainability with minimal energy usage and reduction of energy usage in computing and manufacturing. This definition needs to be expanded.
    - Another consideration in defining sustainability should be the *data center locality.* If one places data centers far away from energy sources, there is a major cost of electricity distribution to data centers. If one places data centers close to energy sources, there might be a major cost of moving data to/from users. Both solutions have their advantages and disadvantages.

**Challenges:** There are several challenges when considering data centers and their energy-aware optimization.

A. E*quity and people* need to be considered. How does one integrate a sustainable data center into a local community and how can it benefit a local community? The answer to these questions could change the evaluation metric of data centers. Instead of only talking about performance, energy, carbon emission of data centers, one could define a metric that accounts for the 'people who are impacted by a data center'. There is a major NIMBY (not in my backyard) problem for large data centers since these data centers get negative community feedback due to noise, increase in local marginal pricing, power grid and power lines, and other concerns. Hence, approaches such as giving back to the community the excess heat, and/or tax credits and revenues could be considered.
B. We need to more broadly quantify the overall environmental impact of data centers. Computing has been optimized and can be energy-efficient, but communication and storage are not efficient. There are trends to estimate energy for computing in AI/Machine Learning Accelerators, supercomputers, and compute-intensive applications (https://arxiv.org/abs/2210.17331). There are energy estimations across layers of computing for large scale ML applications, NLP (Natural Language Processing), Scientific Computing, Cryptocurrency Mining

(https://arxiv.org/pdf/2310.07516.pdf), but energy estimations for network and storage layers are missing.

**Approaches and Recommendations**: Data centers are energy inefficient when using energy to power the system, when aiming to take the heat out of the system, and when considering electronics to maintain the system.

Hence, we need new *cost-effective approaches* that include

- Adaptable infrastructure because the technology advances so quickly,
- Inclusion of green building and environmental planners,
- Quantification of impact of improvements before one starts with optimization of data centers,
- Willing business leaders to invest and act on sustainable data centers,
- Knowledgeable data center researchers that understand the optimization and algorithm side as well as the computing system design and architecture side, and
- Knowledgeable IT administration workforce of research data centers so that they can engage in meaningful sustainability dialogs with data center researchers.

## 3.3 Integration of Computing with the Grid

This break-out session discussed three important questions, challenges and potential approaches.

**Questions:** The three questions that one has to consider when considering integration of computing with the grid are:

- Can large scale computing be used to help *control the frequency of the smart grid*?
- If one considers *distributed data center infrastructures* with smaller decentralized data centers such as edge data centers, what data would be processed locally and what could be processed at the data centers?
- As more and more computing IoT (Internet of Things) devices get plugged into the Grid (*e.g.,* electric cars), and the power demand not just from data centers but also from other pervasive computing devices (*e.g.,* IoT in smart buildings) increases, what changes need to happen to the energy Grid(s)?

**Challenges:**

A. The academic community is way *behind the state-of-the-art* (SOTA) in the area of renewable energy use and data centers. At this point, the industry greatly surpassed academic computing in terms of facility size, power use, and scale of renewable energy and computing studies.
B. Because of our Economic System paradigm, many large scale data centers are designed for *overcapacity when considering future (and peak) usage* (*e.g.,* overprovisioning bandwidth, compute power, energy). In the case of energy, why is it not possible to consider paradigms where one would delay training part of AI models during off-peak execution? Another challenge in this space is the lack of available tools that accurately measure carbon usage of AI workloads. It was mentioned that, for example, current off-the-shelf carbon tracking tools for real-world ML research projects have multiple conflicting and inaccurate assumptions.
C. Computing researchers often explore energy aspects of data centers in isolation from Grid experts. Organizations that run large scale power grids are not always interested in talking with researchers regarding energy and carbon emission. However, the break-out session researchers stressed that we need computing experts talking and being able to work with electrical grid operators.

**Approaches and Recommendations**: The break-out session researchers discussed multiple approaches and recommendations.

- *Metrics, measurements, and distributed computing dispatching* are needed if one wants to have a good understanding of carbon emissions for compute loads. With these measurements and information, one can then minimize power usage in the data center.
- *Placement of smaller data centers inside of renewable compounds* was discussed, *i.e.,* placing smaller data centers close to wind farms or solar farms. Such an approach may, however, have remote access and security problems that need to be considered.
- One needs to consider *reuse of waste heat*. An example was brought up from Quebec where the system Qscale [4] reuses waste heat for agriculture and other heat intensive industries.
- *Funding for cross-domain collaboration* between computing and energy sectors is needed.

# 4.0 Computing for Sustainability

As the ["Sustainable Computing" Session](#), the "Computing for Sustainability" session started with a panel that brought together researchers from diverse backgrounds but with a shared interest in exploring how computing could advance solutions to sustainability problems. The panel involved five (5) panelists, [Canek Fuentes Hernandez](#) from Northeastern University, [Rina Ghose](#) from the University of Wisconsin-Milwaukee, [Yannis Ioannidis](#) from the National and Kapodistrian University of Athens and also representing the [Association for Computing Machinery (ACM)](#) in his capacity as its current President, [Rahul Mangharam](#) from the University of Pennsylvania, and [Alfonso Morales](#) from the University of Wisconsin-Madison.

The panelists brought forward different perspectives and topics that reflected their background and expertise, several of which intersected themes highlighted in the ["Sustainable Computing" panel](#) of the previous day. This is not overly surprising given the growing carbon footprint of computing itself, as the first panel expanded upon. Specifically, themes that emerged from the panelists' presentations included:

- The potential benefit of taking guidance from nature, in particular the trophic pyramid[2], in structuring the architecture and organization of our computing ecosystem. Of particular relevance is increasing our awareness of how computing capabilities and access to power relate to each other across the different layers of our computing hierarchy and to the "cradle-to-grave" carbon footprint of our computing ecosystem.
- The growing importance and steadily improving quality of geospatial data available to monitor a wide range of sustainability metrics. Devising computing solutions capable of leveraging this data at scale, both spatially and temporally, is a major opportunity.
- The fact that while computing plays an increasingly important role in sustainability challenges, ensuring that our efforts have a meaningful impact calls for casting them in the broader context of existing societal frameworks such as the [17 United Nations Sustainable Development Goals](#).
- An emphasis on open access broadly construed, *i.e.,* the ability to not just reuse but also verify, as an essential ingredient to long-term success, one that encompasses not only data, algorithms, and devices, but also aspects of maintenance and repair. The latter is of particular relevance when targeting application areas, *e.g.,* agriculture, with operational structures that differ vastly from those of computing.
- The importance of the "human factor", especially when it comes to fostering the adoption of new solutions, especially in populations with limited exposure to computing.
- The opportunities presented by data-driven machine learning approaches to intelligently learn and control the energy consumption of buildings in response to energy availability, while at the same time acknowledging the many challenges posed by the noisy and often unreliable nature of the data that such systems need.

---

[2] A model for interactions in the food chain associated with biological systems, which reflect how food energy passes from one trophic level to the next across the food chain.

The panel concluded with questions from the audience and discussions with the panelists.

As the previous panel, this was then followed by parallel break-out sessions along the themes identified from the response to the RFI (See [Table 1](#)). Two break-out sessions targeted the themes of "Computing & climate modeling" and "Computing for agriculture & food systems", while the other two both focused on the topic of "Computing for smart infrastructure & materials" to account for the greater number of attendees who had expressed interest in that topic.

## 4.1 Computing and Climate Modeling

The break-out sessions devoted to climate modeling took place in-person and online, with the conclusions reported below coming from both sets of break-out sessions.

The main themes that emerged from the discussions can be grouped in three major categories: **(i)** The importance of human factors in climate modeling, **(ii)** The need for openness in both models and the data on which they rely, and **(iii)** The interdisciplinary nature of both problems and solutions and the unique challenges this creates. Specifically, the participants in the break-out sessions converged on the following conclusions and recommendations:

1. **The importance of human factors in climate modeling**

    - The importance of climate modeling as a scientific endeavor notwithstanding, its results need to be interpreted as a function of their impact on human populations. Not everyone handles, let alone perceives, changes the same way, and tolerance to the impact of climate change varies as a function of occupation, income, local resources, etc. It is, therefore, essential to incorporate those human factors, both in the modeling of climate changes and when attempting to interpret the results of those models.
    - A corollary of the heterogeneity in human factors in determining responses to climate changes is that models themselves need to reflect this heterogeneity. In particular, while many models have sought to provide a *global* perspective, often focusing on how large population centers, *e.g.,* mega-cities, might be affected by climate change, it is equally important to develop models capable of capturing the impact of *local* changes. Encouraging a greater focus on models that can account for local factors and predict local impact is an area that requires greater scientific focus. Additionally, such a focus can often lead to more actionable outcomes, since local actions are often easier to implement than global ones.
    - There is much evidence pointing to human activities as one of the causes for climate change, but the cumulative nature of those activities makes it difficult for individuals to effectively gauge how changes in their personal behavior might eventually have a global impact. It is, therefore, essential to develop enhanced means of communicating those outcomes in ways individuals can

understand and relate to. Technologies such as Virtual Reality (VR) or games can be effective in realizing such a goal, but they require close partnerships between computer scientists and communication experts. Initiating and fostering such collaborations is needed, if only as an important tool in communicating the need for and urgency of proposed changes.

2. **The need for openness in both models and the data on which they rely**

    ○   We undeniably have access to an unprecedented amount of data that can be used to predict climate changes and learn how they may respond to different actions.  This situation, however, creates its own challenges.  The magnitude of the storage requirements for such preservation is one of them, but it is compounded by the need to make that data easily accessible to all. Relevant climate data is commonly acquired by different communities that rely on different data formats, indexing mechanisms, curation processes, etc. This often results in data silos that make data sharing across communities difficult.  New efforts for making climate data openly accessible to all are, therefore, needed to facilitate its long-term sharing and unlock the promises of scientific advances that it holds.
    ○   The need for open-access is not limited to data alone, and extends to the models that consume this data.  Open-access is needed not only to facilitate reusability, but more importantly to enable causal inference.  Openness into how models produce their outcome, including how those outcomes may change as inputs and assumptions vary, is often essential to convince stake-holders to take the actions that the models recommend.  Models' openness alone is obviously not always sufficient to demonstrate causality, but it is a necessary component to enabling such inference.

3. **The interdisciplinary nature of both problems and solutions and the unique challenges this creates**

    ○   Just as climate modeling is transdisciplinary science, computing for climate science is transdisciplinary computing. Successful application of computing solutions to climate modeling requires developing a common understanding of application needs and how they best map onto available computing technologies.  This is a process that is common to many application areas that computing can positively impact, but one that takes time and effort.  It is, therefore, essential to continue fostering opportunities like this workshop that facilitate such interactions and the development of common languages and understanding that such collaborations require.  Similarly, interdisciplinary funding programs structured to encourage participation and contributions from across distinct communities are also needed.
    ○   Finally, the deployment of solutions that climate modeling might generate typically require the sustained involvement of the communities that were initially tapped to assist in the development of those solutions. Maintaining

such engagement beyond the standard lifetime of a scientific project calls for mechanisms that are often not available through standard funding mechanisms. It is, therefore, desirable to explore how such longer-term community sustenance efforts might be supported

In addition to the above findings and recommendations, the break-out sessions also echoed themes developed in the Sustainable Computing sessions. They highlighted the need to make the environmental cost of computing (for climate modeling or otherwise) more apparent to computer scientists, as well as making climate models themselves more energy aware and efficient.

## 4.2 Computing for Agriculture and Food Systems

The break-out sessions devoted to agriculture and food systems took place in-person and online, with the conclusions reported below coming from both sets of break-out sessions.

The discussions surfaced a number of problems and questions specifically aimed at computing as a tool to address some of the many challenges that agriculture and food systems face, but they also broadly echoed several of the themes that emerged during the parallel break-out sessions devoted to improving the sustainability of computing itself (see [Section 3.0](#)).

1. **Opportunities and challenges in leveraging computing to develop solutions to agriculture and food systems problems**
    - **Opportunities**
        - There is a tight coupling between climate modeling and agriculture, with both in a position to help the other (this was also echoed in the [Climate Modeling](#) break-out sessions):
            (i) How can computing help food production become a carbon consumer rather than a carbon producer? For example, models of crop growth and carbon consumption as a function of soil type and climate could be used to promote crop selection and rotation to maximize carbon absorption.
            (ii) How can computing assist food production reduce its own carbon footprint? This spans sensing and (AI/ML) models to, for example, predict crop response as a function of weather and soil conditions towards minimizing water and fertilizer usage, or optimize the logistics of food distribution to minimize its carbon footprint and food wastage, or providing early detection of illness in both crops and animals.
        - Can we develop community based data-sharing tools for coordinating agricultural work to increase yields and improve crop selection? This should not only be aimed at addressing challenges that stem from the heterogeneity of the available data, but should also address user interface challenges. A possible example could

be in the form of a "dashboard" for farmers and farm workers to provide not only visualization but also communal insights into difficult problems such as predicting yields, climate models, shipping perishable goods, diversifying soil usage, etc. This should span different time-scales, including 6 months or longer, to match the different needs of agriculture.
- Is it possible to devise a digital farm assistant, *e.g.,* a "FarmGPT" to provide individualized guidance on matters such as seed selection, fertilizer choices as a function of soil type, location, budget, etc. This would require access to personalized data to ensure locally relevant recommendations, training on data from a wide range of sources to accommodate heterogeneity in environments and users, and the development of a user interface adapted to farmers and capable of gaining their trust.

- **Challenges (addressing those represents another set of opportunities)**
  - Data privacy and ownership. Having farmers reveal individual data may expose them to possible liabilities, *e.g.,* water usage data, and this creates an inherent tension between the potential benefits of precision agriculture and the data privacy risks it forces farmers to incur. Can we anonymize individual farmer's data while preserving its usefulness? This appears challenging especially due to the geospatial nature of much of the data. A possibly useful parallel is that of the health system where regulations such as HIPAA were drafted to ensure data privacy while enabling data sharing.
  - Data heterogeneity in both representation and how it is shared and disseminated. There are no common and agreed upon agriculture and food data ontologies, and the mechanisms and repositories used for data sharing are disparate and lack any form of integration. This is especially so for data collected by local governments that is often kept in silos in spite of its potential usefulness.
  - Heterogeneity in stakeholders' population and how to best engage them in using computing-based solutions. Gaining the trust of such a diverse community requires a focused effort and the involvement of community and workforce development experts[3]. Further, sustaining this trust calls for demonstrating the practical benefits of predictive, data-oriented solutions, *e.g.,* in making land management decisions oriented that improve future farm resilience.

2. **Connections with Sustainable Computing**

---

[3] The https://www.climatehubs.usda.gov/hubs/midwest/tools/useful-useable-u2u highlights approaches that worked effectively to engage farmers and other stakeholders in using climate information for decision making, and provides links to a set of useful tools.

- Can bio-materials provide more sustainable solutions to the manufacturing of computing systems?  This extends to sensors that might bio-degrade in the environment once they cease to be operational, to bio-materials capable of replacing silicon as the substrate on which computing systems are built.
- Given the importance of water in agriculture and food production, how can we ensure a more harmonious coexistence between computing facilities and food production when it comes to their respective water usage.

## 4.3 Computing for Smart Buildings and Materials

The in-person and remote break-out sessions used slightly different formats to surface and discuss relevant themes and challenges, but their conclusions were broadly consistent and can be grouped into three major thrusts as listed below:

1. **Existing vs. new constructions**.  There is a need to broaden the scope of the original question to focus not only on new constructions, but also on operational costs.
    - The majority of buildings already exist and most remain in operation for 50+ years.  Current "standards" focus mostly on pre-occupancy design metrics with no post-occupancy validation.  A greater focus should be given to post-occupancy efficiency, if only to validate the expectations of the initial design.
    - A corollary of the predominance of existing buildings is the need to focus on developing low-cost retrofitting solutions. Renovations are expensive and can themselves have a high environmental cost.  Devising efficient renovation approaches could, therefore, yield significant environmental benefits.
2. **Optimization Complexity**.  Controlling the energy consumption of a building is a multi-faceted and complex problem.
    - It is important to be aware of the relative scope for optimization.  40% of the energy consumption of a building occurs while it is unoccupied, *i.e.,* relatively static and with limited opportunities for adjustments and optimization.  In addition, energy consumption patterns exhibit a long tail and there is limited flexibility (around 10%) in being able to control overall consumption.
    - Energy price volatility is a significant factor behind energy optimization, and it introduces a temporal dimension that adds significant complexity.  In other words, we need energy flexibility just as much as energy efficiency.
    - Each building is unique, which makes it hard to develop general solutions. In addition, building control systems (software and hardware as well as the data they generate) are usually closed/proprietary, and the data they produce is often "dirty", both on the design front (missing data) and the operational front (faulty sensors).
3. **Human, social, and policy factors**. Control systems need to interact with humans, which introduces unpredictability as well as diversity in what might be deemed "optimal" policies.  Additionally, buildings are part of a broader ecosystem that is the subject of numerous regulations that need to be taken into account.
    - Different types of occupations translate in different levels of tolerance to variations in environmental conditions.  Cultural norms can also have a similar effect, as do factors such as age and social and economical standings.  Systems need to be able to learn and adapt to such differences.

- Conversely, humans themselves adapt and react to changes, which further complicates attempts at optimization.
    - Aspects of data privacy should also be considered, especially in settings where individual buildings interact as part of a broader optimization setting, *i.e.,* city-wide or throughout a region.
    - While (energy) efficiency and cost are important factors, it is also important to ensure that solutions do not contribute to creating greater inequities across occupants of different buildings. Policy and economic factors must, therefore, be accounted for when devising solutions.
    - CISE researchers have limited domain expertise when it comes to buildings and their operation. Conversely, building experts are usually not trained to generate high quality data sets (starting from hardware design, sensors that reduce data waste), or organize and merge databases. Both affect our ability to develop effective solutions
    - Can buildings be leveraged to monitor the humans that occupy them and assist with improving well-being and safety?

Several representative problems associated with the above challenges emerged during the discussions, and are listed next as representative examples:

- Can we develop efficient learning-enabled systems that automatically acquire/learn information about a building and use it to develop scheduling and control policies?
    - Can this be extended to a human-centric approach to sustainable buildings, one that would learn behavioral and cultural differences, and offer individualized incentivization.
- Can we incentivize energy storage at the local level to facilitate energy allocation and control at different time-scales?
    - Local energy storage can make for more efficient policies by mitigating temporal variability, but it comes at a cost, and justifying this cost requires commensurate incentives. Developing those incentives is just as much a policy and business problem as it is a technical one.
- Ontologies, datasets, and real-time Building Information Modeling (BIM) tools are needed:
    - The multiplicity of proprietary standards and the lack of general methodologies to account for missing or erroneous data is a major impairment to the development of effective solutions.
    - Can we anonymize building data and interactions with the grid to address privacy concerns?
- Can we extend building control to focus not just on energy consumption but also on water consumption and air quality?
- Developing easily customizable digital twin building models could go a long way towards facilitating the development and testing of control policies.
    - Incorporating the human component into such models may, however, be a challenge.

# 5.0 Summary and Recommendations

This section identifies potential recommendations (to NSF and other funding agencies) towards realizing some of the opportunities the workshop helped reveal. Recommendations are split across the themes of "Sustainable Computing" and "Computing for Sustainability", with a few that cut across both themes pulled out in a subsection of their own.

## 5.1 General Recommendations

**[G1].   Promoting the development of open data models and sustaining repositories that support privacy mechanisms and facilitate data sharing.**

This was a recurrent theme across the sessions, as *open* access to data is essential to any solution targeting problems identified in both the "Sustainable Computing" and "Computing for Sustainability" break-out sessions. Access to data is, however, a multi-faceted problem spanning what, where, and how.

(i). It starts with the specification of ontologies and data formats to ensure a common understanding of what the data captures and the relationships that exist across data. This is lacking in many domains, not just in application domains that may not be traditionally "data rich", but also in computing itself where, for example, data on sustainability metrics are often unavailable or fragmented.
- In particular, there is a need for sustainability-specific metrics and validation services performed over open data models to measure and quantify the sustainability of computing; in the process enabling design and policy choices. Data is needed beyond energy measurements of data centers and their carbon emission, but also to capture the impact of e-waste on people, the environment, and other side-effects.

(ii). The definition and specification of data and its format is, however, only the beginning. This must be mapped to mechanisms or systems capable of reliably generating the required data. This remains a priority, as while there has been enormous progress in our ability to sense and monitor both the environment and systems, many "blind spots" remain.

(iii). Creating public repositories with the goal of long-term data preservation and accessibility and with clear policies for data access and sharing, including securely and without jeopardizing the privacy of data sources, is another essential step in realizing the kind of ubiquitous data access that is needed.

(iv). Finally, there is a need to encourage the development and maintenance of reusable tools (across communities) for data integration and sharing, including with the ability to verify data provenance while preserving privacy, both of which may be at odds with each other.

**None of the above efforts readily map into traditional scientific endeavors, especially aspects of long-term support and maintenance, but their importance to successful science will only grow with the role of data itself. It is, therefore, essential to develop funding mechanisms that support both initiating and sustaining such efforts.**

**[G2]. On the importance of accounting for human and social dimensions in most sustainability problems.**

This represented another theme that surfaced across many of the discussions, although it manifested itself through a wide range of instantiations, each possibly worthy of a separate recommendation.

Human factors obviously affect climate change, with many arguing that they are partially responsible for it, but they are equally important when seeking to interpret the predicted impact of climate change, or the likely adoption of proposed mitigation measures. Similar arguments hold when considering the role of human factors in the control loop that determines the energy efficiency of buildings, or in how they influence the usage and purchasing decisions that affect the lifecycle of computing devices.

More broadly, humans are often behind the policy decisions that control the success of proposed solutions, whether they are aimed at improving the energy efficiency of computing or produced by computing to optimize a specific outcome, *e.g.,* improve crop yield. As a result, it is important to develop approaches that improve human understanding of sustainability challenges and/or the benefits of solutions proposed to address them. This encompasses both aspects of visualization as well as interfaces that facilitate individualized interactions.

In summary, sustainability is more than a technical problem. It is inherently coupled to human behaviors and perception. As a result, **it is imperative that initiatives that seek to tackle sustainability, be it the sustainability of computing or how computing can help sustainability itself, be explicit and intentional in how they expect to account for human factors or leverage human factors to achieve their goals.**

**[G3]. The need for cross-disciplinary perspectives that ensure a common understanding of not just goals, but also constraints and associated trade-offs.**

While access to data and metrics that quantify the sustainability cost of computing is essential (recommendation **[G1]**), it is by itself not sufficient to the realization of sustainable solutions. Computing is a means to an end, *i.e.,* a tool for an application that relies on computing to generate a result. Understanding the needs of that application is a critical step in identifying suitable design trade-offs between sustainability and other relevant criteria such as performance, cost, etc. In other words, sustainable computing solutions need to be aware of how sustainability choices affect the applications that rely on them.

This is in itself a challenge, as solutions need a "cross-layer" perspective that accounts for hardware, software, and application dependencies. This requires not only suitable metrics to measure the effectiveness of solutions at each layer, but also the creation of *global* criteria that capture a holistic view of the impact of design choices across layers. This in turn calls for a cross-disciplinary perspective and dialog across stake-holders.

**It is, therefore, essential to develop best practices that can inform how to realize such cross-disciplinary perspectives. Pilot programs that take a target application through a "from cradle to grave" design cycle might help develop suitable templates.**

## 5.2 Sustainable Computing - Recommendations

> **[SC1]. Cross-Disciplinary Funding to Bring Together Materials, Manufacturing, Computing, Energy Experts**

Computing research is predominantly funded by the NSF CISE directorate and other computing-oriented divisions of other funding agencies. However, if we want to minimize carbon emission and achieve sustainable computing, if we want to achieve a holistic cycle of data centers and power grid, if we want to prolong lifetime of consumer electronics, we need to look at the manufacturing, supply chain and the operational cycle from inception to end of life of computing devices. In particular, while modularity can greatly improve the sustainability of computing devices, developing and more importantly incentivizing the necessary end-to-end ecosystem (from manufacturing to maintenance to recycling) remains a major hurdle. It is as much a technical problem as an economic and logistics one. A similar dilemma arises on the software side from copyright and security issues. Here again, the challenges are not just technical but span legal and economic aspects.

**This means that there is a need for very strong collaborations among the various disciplines involved in the entire lifecycle of computing components. This requires cross-disciplinary research and funding across different directorates (*e.g.,* CISE and Engineering for manufacturing and supply chain carbon emissions of computing elements) and other agencies (*e.g.,* NSF CISE and DOE to work towards a holistic framework and integration between computing and Grid). This should not be limited to large, flagship initiatives, but also extend to smaller-scale seed efforts.**

> **[SC2]. Successful impact calls for focusing on Technology Transfer Efforts and Increased Collaborations between Industry and Academia, and Across Diverse Government Agencies.**

Sustainable computing is an important area for industry, but, as the workshop made abundantly clear, it is not only highly interdisciplinary, it also involves a large number of decision makers with different perspectives and objectives. The development of sustainable computing solutions, therefore, requires programs that, on the one hand, foster greater collaboration between industry and academia, and, on the other hand, target deliverables that can serve as forcing functions to the successful delivery of operational solutions.

For example, better integration of data centers with the Smart Grid, calls for programs that not only bring together computer scientists and energy experts, but also that can sustain those efforts well beyond the design, testing, and validation of an initial solution. This is somewhat of a departure from the traditional funding models on which the NSF CISE directorate and other computing-oriented divisions of other funding agencies rely, but aligns well with the mandate of the recently created TIP directorate. Solutions need to look at the full cycle starting with the inception and birth of sustainability ideas, and extending into the technology transfer of sustainability-related solutions. This requires investments that provide long-term support for personnel, *e.g.,* from research scientists, to graduate students, to full-time programmers, across the different phases of projects. This in turn will necessitate coordination between different funding agencies and stake-holders to ensure the necessary continuity of support.

**It is, therefore, essential to establish collaborative mechanisms and venues that will not only enable increased collaborations between industry, government and academia, but also facilitate continuity of support for those projects across different agencies or directorates (*e.g.,* CISE and TIP for technology transfer).**

> [SC3]. Elevate Sustainability's Profile as a Stand-Alone Topic on Both the
>         Educational and Research Fronts

Unlike the research front that has seen growing activities, on the educational front, academic institutions currently do little to educate the future workforce in sustainable computing. Given the growing needs for such a workforce, we need to expand our undergraduate and graduate curricula to include courses that train future IT administrators and hardware and software engineers in sustainable computing.

Conversely, while there is indeed a growing body of work focused on reducing energy and carbon emission from operating data centers, there is much less research targeting embodied carbon emissions of computing devices, modularity and lifecycle of computing devices, and, on the operational front, how to better integrate data centers with the electric Grid. Consequently, it is important to encourage more work on those aspects of sustainability. One aspect that may be affecting such activities is the fact that, like many other interdisciplinary topics, research on sustainable computing is often not well-recognized in academic institutions and viewed as mostly an "engineering" topic that resides primarily in industry.

**Addressing those issues calls for a multi-prong approach, potentially including the following activities suggested by the workshop participants:**
- **Initiate a curriculum development effort focused on creating and sharing undergraduate and graduate courses in sustainable computing. Additionally,** **consider specialization tracks and certifications in sustainable computing to develop a pipeline of students with the necessary knowledge and skills.**
- **Organize sustainable computing focussed workshops and conferences and/or include sessions in existing venues about sustainable computing.**
- **Leverage professional societies and forums to promote recognition of sustainable computing as a major research track by the academic community and leadership.**
- **Encourage community engagement in developing policies, governance and standardization of sustainable computing.**

## 5.3 Computing for Sustainability - Recommendations

**[C4S1].      Climate models that span from the local to the global.**

The many recent advances in developing more powerful and more accurate climate models notwithstanding, there is a need for approaches that can not only capture global trends, but that are also capable of offering local predictions and recommendations.  The latter are of particular interest in supporting local decisions when assessing the effectiveness of proposed mitigation options.  In general, while climate change is a global problem, its impact varies at the local level, and more importantly, actionable outcomes are often local, *e.g.,* mitigation solutions or identifying vulnerabilities.

**Significant value and effectiveness can, therefore, be unlocked by developing climate models that translate global findings into a local scale and vice-versa. Initiatives aimed at exploring models and systems that capture those interactions should, therefore, be encouraged, including the required community building initiatives.  Sustaining the latter over the long-term may require new funding models.**

**[C4S2].      Sustainable agriculture as a climate strategy.**

Food production is heavily dependent on climate, but at the same time, the scale of food production is such that it can itself have a powerful impact on climate.  It is, therefore, of interest to consider food production and climate models as coupled rather than separate problems.  This means not only encouraging the development of joint models that account for this coupling and explore its impact and consequences, but also building bridges across the different communities (climate, agriculture, computing) that need to contribute their respective expertise.

**Exploring and leveraging opportunities at the intersection of climate and agriculture requires establishing interdisciplinary programs explicitly intended to bring together climate, agriculture, and computing scientists and practitioners, with the goal of creating compelling use-cases for food producers.**

**[C4S3].      Leveraging learning strategies to tackle sustainability problems in the construction and building control area.**

The power of machine learning solutions is by now well-understood and has been demonstrated on a wide range of sustainability problems, *e.g.,* DeepMind GraphCast model for weather prediction [2], or Caltech's climate machine[4]. There are, however, many other areas where their potential is only starting to be explored even if major challenges remain.

Construction and building control are of particular interest. Performance metrics and measurement data are abundantly available even if their quality is uneven and their format often of a proprietary nature. Furthermore, the control and planning problems that arise are inherently complex, often exceeding the capabilities of traditional optimization methods or requiring expensive customized solutions for each individual building. Learning solutions, therefore, represent a promising option, albeit one that is not without challenges. Those include complex spatial and temporal interactions, significant diversity in both operating conditions and desired objectives, and the need for robustness in handling missing or erroneous data.

Those challenges notwithstanding, the potential gain from improving buildings' energy efficiency is significant (in 2023, the combined end-use energy consumption by the residential and commercial sectors accounted for nearly 30% of total U.S. end-use energy consumption[5] [3]).

**Programs that bring together building engineers and computer scientists with the goal of developing learning solutions for more efficient building control should, therefore, be encouraged.**

---

[4] https://clima.caltech.edu/
[5] https://www.eia.gov/tools/faqs/faq.php?id=86

# Acknowledgments

This workshop was supported by NSF grants CCF 2334853/4/5.

The steering committee of the NSG WSCS 2024 workshop was comprised of the following individuals:

- David Allen - University of Texas Austin, Texas
- Annammalai Annammalai - Prairie View A&M University, Texas
- Andrew Chien - University of Chicago, Illinois
- Ian Foster - University of Chicago/Argonne National Lab, Illinois
- Tamar Eilam - IBM Research, New York
- Chandra Krintz - University of California Santa Barbara, California

The material from [Section 2.0](#) was contributed by Anne Mook and Jeni Cross who led the workshop's "*team science"* session.

The following individuals contributed to the editing and refining of the final version of this report.

- Canek Fuentes Hernandez
- Rina Ghose
- Roch Guerin
- Chandra Krintz
- Amy McGovern
- Alfonso Morales
- Klara Nahrstedt

The following individuals served as **panelists** in the panels on "Sustainable Computing" and "Computing for Sustainability".

- Andrew Chien
- Tamar Eilam
- Canek Fuentes Hernandez
- Rina Ghose
- Yannis Ioannidis
- Akshaya Jha
- Kieran Levin
- Rahul Mangharam
- Alfonso Morales
- Christine Ortiz
- Carole-Jean Wu

The following individuals served as **moderators** for the workshop's break-out sessions and helped generate summaries of their discussions. These summaries provided a foundation for the workshop's report.

- **In-person break-out sessions**:
  - Andrew Chien
  - Tamar Eilam
  - Canek Fuentes Hernandez
  - Yannis Ioannidis
  - Akshaya Jha
  - Kieran Levin
  - Rahul Mangharam
  - Alfonso Morales
- **Remote break-out sessions**
  - Ashwin Ashok
  - Aaron Jezghani
  - Sheldon Liang
  - Mike Liebhold
  - Reid Lifset
  - Tania Lorido
  - Jasmine Lu
  - Pritish Parida
  - Shaolei Ren
  - Stefan Robila
  - Alan Sill
  - Raja Sengupta
  - Carol Song
  - James Wilgenbusch
  - Kristin Williams
  - Mengxin Zheng

The following individuals served as **scribes** during the workshop's break-out sessions. Their notes captured the essence of the discussions that took place during those sessions, and were an essential input to the workshop's report.

- **In-person break-out sessions**:
  - Daniel Aliaga
  - Iraklis Anagnostopoulos
  - Andreas Andreou
  - Suman Banerjee
  - Tom Boellstorff
  - Jorge Celis
  - Abhishek Chandra
  - Vidya Chhabria
  - Ayse Coskun
  - Gabe Fierro
  - Nancy Fulda
  - Rong Ge
  - Udit Gupta

- - - Can Hankendi
    - Kurtis Heimerl
    - Daniel Howard
    - Vikram Iyer
    - Benjamin Lee
    - Stephen Lee
    - Nicolas Martin
    - Dejan Milutinovic
    - Ella Neumann
    - Sreepathi Pai
    - Eve Schooler
    - Matthew Sinclair
    - Pingbo Tang
    - Christof Teuscher
    - Christopher Yeh
    - Wangda Zuo
- **Remote break-out sessions**
    - Charalampos Chelmis
    - Catherine Gill
    - Sathish Gopalakrishnan
    - Kathryn Kelley
    - Michael Krasowski
    - Sheldon Liang
    - Qian Lou
    - Pratibha Raghunandan
    - Esther Roorda
    - Sumit Sen
    - Brandon Tran

The following individuals served as workshop website and EDAS chairs:

- Ryan Zhang
- Bo Chen

# Appendices

## A.1 List of Questions to Panelists and Attendees

Those questions were sent to both panelists and attendees ahead of the workshop and were aimed at kick-starting and focusing the discussions.

**Sustainable Computing**

1. Datacenters, Energy and Optimization

    - What are key obstacles, barriers and challenges, technical or otherwise, to ensuring the sustainability of datacenters and minimizing their energy footprint and carbon emission while still meeting performance and reliability goals?
        i. This encompasses technologies, operational aspects, and application behavior
    - What tools and information are needed to overcome those obstacles?
    - What business practices need to change or economic incentives created to make solutions practical and foster adoption?
    - Other aspects of relevance beyond those listed above?

2. Integration of Computing with the (Smart) Grid

    - What are key obstacles, barriers and challenges, technical or otherwise, to ensuring mutually beneficial interactions between computing systems and a smart grid that is rapidly evolving, *e.g.,* seeing an increasing prevalence of renewable sources and a shift from a centralized architecture to a more decentralized one with many interconnected microgrids?
        i. This encompasses both computing as a user of the grid, and computing as an enabler for a smarter grid
    - What tools and information are needed to overcome those obstacles?
    - What business practices need to change or economic incentives created to make solutions practical and foster adoption?
    - Other aspects of relevance beyond those listed above?

3. Modularity and Lifecycle of Computing

    - What are key obstacles, barriers and challenges, technical or otherwise, to improving the sustainability of computing systems, including aspects of fabrication, operation, and maintenance?
        i. This encompasses aspects of upgradeability and evolving workload requirements
    - What tools and information are needed to overcome those obstacles?
    - What business practices need to change or economic incentives created to make solutions practical and foster adoption?
    - Other aspects of relevance beyond those listed above?

**Computing for Sustainability**

1. Computing and Climate Modeling

- What are key obstacles, barriers and challenges, technical or otherwise, to developing computing solutions that can more accurately model/predict climate changes and/or the impact that various remediation solutions may have?
    i. This encompasses scaling and performance aspects, as well as dealing with heterogeneous data sources, and persistence of information at different time-scales.
- What tools and information are needed to overcome those obstacles?
- What business practices need to change or economic incentives created to make solutions practical and foster adoption?
- Other aspects of relevance beyond those listed above?

2. Computing and Agriculture and Food Systems

    - What are key obstacles, barriers and challenges, technical or otherwise, to deploying and maintaining computing solutions that enable high-performance agriculture and efficient food distribution?
        i. This encompasses deployment aspects and in-situ sustainability across environments with different cost and skills structures, and equipment with vastly different life cycles.
    - What tools and information are needed to overcome those obstacles?
    - What business practices need to change or economic incentives created to make solutions practical and foster adoption?
    - Other aspects of relevance beyond those listed above?

3. Computing and Smart Building

    - What are key obstacles, barriers and challenges, technical or otherwise, to developing computing solutions that can support more efficient buildings' construction and operation?
        i. This encompasses materials and processes involved in the creation and maintenance of buildings, and the tools for monitoring the status of the built environment and operating it in a sustainable manner..
    - What tools and information are needed to overcome those obstacles?
    - What business practices need to change or economic incentives created to make solutions practical and foster adoption?
    - Other aspects of relevance beyond those listed above?

## A.2  List of Relevant Programs (as of April 2024)

**NSF [Design for Environmental Sustainability in Computing](#) (DESC)**

The goal of the DESC program is to address the substantial environmental impacts that computing has through its entire lifecycle from design and manufacturing, through deployment into operation, and finally into reuse, recycling, and disposal.

**NSF [Civic Innovation Challenge](#) (CIVIC)**

CIVIC is a research and action competition that accelerates the transition to practice of foundational research and emerging technologies into communities through civic-engaged research.

**NSF [Smart and Connected Communities](#) (S&CC)**

The goal of the S&CC program solicitation is to accelerate the creation of the scientific and engineering foundations that will enable smart and connected communities to bring about new levels of economic opportunity and growth, safety and security, health and wellness, accessibility and inclusivity, and overall quality of life.

**NSF 24-058: Dear Colleague Letter**: [Supporting Computing & Networking Research for a National Discovery Cloud for Climate](#) (NDC-C)

This DCL from the CISE Directorate encourages the research community to submit proposals to the Computer Systems Research (CSR) program or the Networking Technology and Systems (NeTS) program in support of the creation or enhancement of a National Discovery Cloud for Climate (NDC-C).

**NSF 24-022: Dear Colleague Letter: [Build a Resilient Planet](#)**

This DCL from the Directorate of Geosciences announced a series of Dear Colleague Letters (DCLs) highlighting priority research areas and encouraging submission of proposals in several areas of interest. Links to these DCLs can be found at the NSF/GEO [web page](#).

**NSF [Collaborations in Artificial Intelligence and Geosciences](#) (CAIG)**

The CAIG program seeks to advance the development and adoption of innovative artificial intelligence (AI) methods to increase scientific understanding of the Earth system and enable significant breakthroughs in addressing geoscience research question(s).

**NSF 24-045: [Dear Colleague Letter: Funding Opportunities for Engineering Research to Achieve Net-Zero Climate Goals by 2050](#)**

This DCL from the Directorate for Engineering encourages the submission of research and education proposals related to Net-Zero Climate Goals, including innovations to create a Circular Economy.

**NSF [Regional Innovation Engines](#)**

The NSF Engines program — led by the NSF Directorate for Technology, Innovation and Partnerships (TIP) — envisions supporting multiple flourishing regional innovation ecosystems across the U.S., spurring economic growth in regions that have not fully participated in the technology boom of the past few decades.

**NSF [Plant Genome Research Program](#) (PGRP)**

The PGRP encourages the development of innovative tools, technologies, and resources that empower a broad plant research community to answer scientific questions on a genome-wide scale.

**NSF [Biodiversity on a Changing Planet](#) (BoCP)**

The BoCP program encourages proposals that integrate ecological and evolutionary approaches in the context of the continual gain, loss, and reorganization of biodiversity on a changing planet.

**NSF [Organismal Response to Climate Change](#) (ORCC)**

The ORCC calls for proposals that integrate the study of organismal mechanisms of response to climate change (ORCC) with eco-evolutionary approaches to better predict and mitigate the effects of a rapidly changing climate on earth's living systems.

**NSF [Convergence Accelerator Phases 1 and 2 for the 2023 Cohort](#) - Tracks K, L, M**

The Convergence Accelerator program seeks to transition basic research and discovery into practice—to solve high-impact societal challenges aligned with specific research themes (tracks). The 2023 cohort includes the following: Track K: Equitable Water Solutions, Track L: Real-World Chemical Sensing Applications, and Track M: Bio-Inspired Design Innovations.

**NSF [Algorithms for Modern Power Systems](#) (AMPS)**

The AMPS program will support research projects to develop the next generation of mathematical and statistical algorithms for improvement of the security, reliability, and efficiency of the modern power grid.

**DOE [Electronics Scrap Recycling Advancement Prize](#) (E-SCRAP)**

The U.S. Department of Energy (DOE) Electronics Scrap Recycling Advancement Prize three-phase competition will award up to $4 million to competitors to substantially increase the production and use of critical materials recovered from electronic scrap—or e-scrap.

**DOE [Grid Resilience and Innovation Partnerships](#) (GRIP)**

The GRIP program aims to enhance grid flexibility and improve the resilience of the power system against growing threats of extreme weather and climate change.

**IEDO FY24 [Energy and Emissions Intensive Industries](#) FOA**

This funding opportunity announcement (FOA) from DOE Industrial Efficiency and Decarbonization Office (IEDO) focuses on applied research, development, and demonstration (RD&D) for the highest GHG-emitting industrial subsectors, specifically:

chemicals and fuels; iron and steel; food and beverage; building and infrastructure materials (including cement and concrete, asphalt pavements, and glass); and forest products.

**USDA FY 24 [Open Data Framework](Open Data Framework)**

The Open Data Framework program will build a framework needed to create a neutral and secure data repository and cooperative where producers, universities and nonprofit entities can store and share data in ways that will foster agricultural innovation and will support technological progress, production efficiencies, and environmental stewardship.

## A.3 List of Publication Venues

The Sustainable Computing for Sustainability research topics can be published in various venues as follows:

**Sustainable Computing Venues:**

Journals/Magazines:

- [Artificial Intelligence for the Earth Systems](#)
- [IEEE Transactions on Sustainable Computing](#)
- [IEEE Pervasive Computing Magazine](#)
- [IEEE Journal on Modern Power Systems and Clean Energy](#)
- [IEEE Transactions on Smart Grid](#)
- [IEEE Transactions on Sustainable Energy](#)
- [Springer Energy Informatics](#)
- [Green Software Foundation](#)

Conferences:

- [ACM International Conference on Future and Sustainable Energy Systems (ACM e-Energy)](#)
- [ACM CPS-IoT Week](#)
- [ACM/IEEE International Conference on Internet of Things Design and Implementation (ACM/IEEE IoTDI)](#)
- [ACM SIGCAS Conference on Computing and Sustainable Societies (COMPASS)](#)
- [IEEE International Conference on Pervasive Computing and Communications (IEEE PerCom)](#)
- [IEEE International Conference on Communications, Control, and Computing Technologies for Smart Grids (IEEE SmartGridComm)](#)
- [International Green and sustainable computing Conference](#)
- [IEEE International Conference on Sustainable Computing and Communications (IEEE SustainCom)](#)
- [ICLR Workshops: ICLR 2024 Workshop: Tackling Climate Change with Machine Learning](#) (these workshops rotate among NeurIPS and ICLR and ICML)

- [AMS annual meeting](#)

- Annual AMS AI conference (this is the one at the 2025 AMS annual meeting) [24th Conference on Artificial Intelligence for Environmental Science](#)

- [AGU annual meeting](#)

**Computing for Sustainability Venues:**

Climate and Environment:

- [Climate Change AI](#)
- [Elsevier Current Research in Environmental Sustainability](#)
- [Elsevier Climate Change Ecology](#)

- [Elsevier Climate Risk Management](#)
- [Elsevier International Journal on Climate Change Strategies and Management](#)
- [Elsevier Urban Climate](#)
- [Elsevier Sustainability Analytics and Modeling](#)
- [Springer Environmental Systems Research](#)
- [MDPI Climate](#)
- [MDPI Sustainability](#)
- [Climate Informatics Conference](#)
- [AMS (American Meteorological Society) Annual Meeting](#)
- [AMS Symposium on High Performance Computing for Weather, Water, and Climate](#)

Agriculture:

- [Elsevier Computers and Electronics in Agriculture](#)
- [Wiley Journal of Field Robotics](#)
- [Elsevier Climate Smart Agriculture](#)
- [Elsevier Journal of Agriculture and Food Research](#)
- [Robotics: Science and Systems (RSS) Conference](#)
- [MDPI Agriculture](#)
- [Center for Digital Agriculture Conference](#)
- [USDA Agricultural Outlook Forum](#)

Smart Buildings:

- [Elsevier Green Technologies and Sustainability](#)
- [ASME Journal of Engineering for Sustainable Buildings and Cities](#)
- [CONF.ITECH: International Conference on Technological Imagination in the Green and Digital Transition](#)
- [MDPI Buildings](#)
- [ACM International Conference on Systems for Energy-Efficient Buildings, Cities, and Transportation (BuildSys)](#)